\documentclass[letterpaper]{article}
\usepackage[preprint]{preprint}
\usepackage[hyphens]{url}
\usepackage{graphicx}
\usepackage{natbib}
\usepackage{caption}
\usepackage{hyperref}
\hypersetup{colorlinks=true, urlcolor=blue, citecolor=black, linkcolor=black, breaklinks=true}
\usepackage{algorithm}
\usepackage{algorithmic}
\usepackage{amsmath}
\usepackage{amssymb}
\usepackage{booktabs}
\usepackage{multirow}
\usepackage[table]{xcolor}   

\definecolor{rankone}{HTML}{FBC7C6}
\definecolor{ranktwo}{HTML}{FBDCC0}
\definecolor{rankthree}{HTML}{FBF3C6}
\newcommand{\ca}{\cellcolor{rankone}\bfseries}
\newcommand{\cb}{\cellcolor{ranktwo}\bfseries}
\newcommand{\cc}{\cellcolor{rankthree}\bfseries}

\title{QuARC-GS: Quantized Anchored Residual Coding for Compact Dynamic Scene Streaming with Gaussian Splatting}

\author{
    Vu Trung Nghia Nguyen\equalcontrib,
    Yuchen Wang\equalcontrib,
    Kyung Chul Lee,
    Kevin C. Zhou\corresponding
}
\affiliations{
    University of Michigan\\
    \{nvtnghia, wangyuch, kcleebme, kczhou\}@umich.edu
}

\begin{document}
\maketitle

\begin{abstract}
3D scene representation techniques such as neural radiance fields (NeRFs) and Gaussian splatting have made substantial progress in novel view synthesis, achieving high-quality renderings from arbitrary view angles. More recently, such techniques have been extended to dynamic 3D scenes; however, achieving sustainable online free-viewpoint video (FVV) streaming remains challenging, especially for longer videos, due to significant storage demands of detailed scene representations and high reconstruction/rendering speed needs. To address these challenges, we propose Quantized Anchored Residual Coding Gaussian Streaming (QuARC-GS), a quantization-aware 4D scene optimization framework for online dynamic scene reconstruction that achieves ultra-high compression while maintaining reconstruction speed and quality. QuARC-GS represents a scene using a single canonical frame and highly compressed per-frame residuals. Specifically, we compress each residual through two complementary strategies targeting motion, appearance, and densification. We introduce quantization-aware anchor deformation, which suppresses insignificant motion updates while preserving meaningful deformations, maintaining reconstruction quality under low-storage streaming. Furthermore, we design a change-gated densification strategy that allocates new Gaussians only in regions exhibiting genuine temporal changes, effectively eliminating redundant appearance updates and reducing storage overhead. Extensive experiments on widely used datasets demonstrate that QuARC-GS enables competitive reconstruction quality and training speed while cutting per-frame storage by up to 11$\times$ compared to the state-of-the-art. 
\end{abstract}

\begin{links}
    \link{Code}{https://github.com/high-performance-computational-optics/QuARC-GS}
\end{links}
\section{Introduction}
\label{sec:intro}

\begin{figure}[t]
    \centering
    \includegraphics[width=\linewidth]{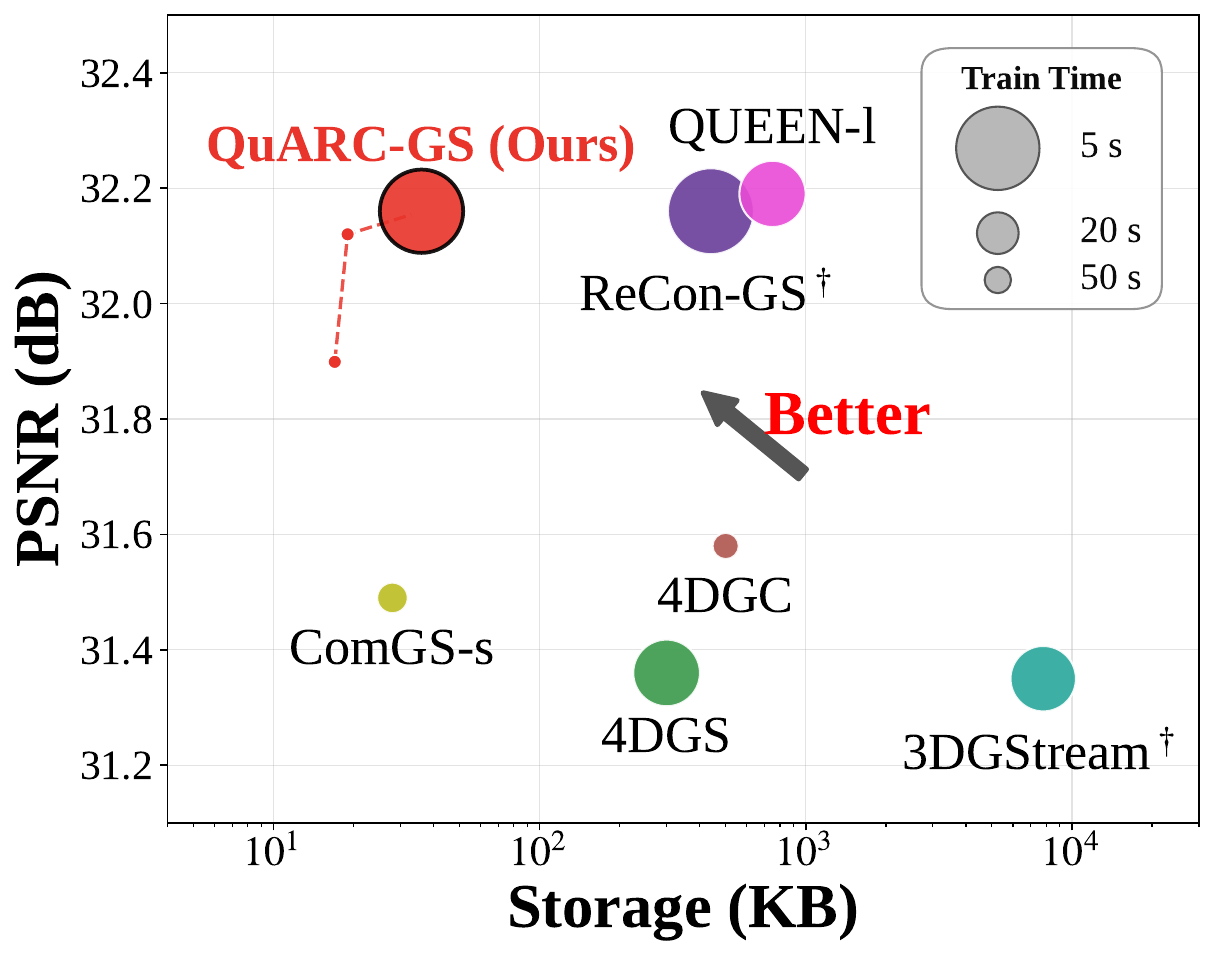}
    \caption{Rate-distortion comparison on N3DV \citep{li2022neural3dvideo}. QuARC-GS streams dynamic scenes at a
    fraction of the per-frame storage of recent online methods
    (ReCon-GS$\dagger$~\citep{fu2025recongs}, QUEEN~\citep{girish2024queen},
    ComGS~\citep{chen2025comgs}, 4DGC~\citep{hu2025_4dgc},
    3DGStream$\dagger$~\citep{sun2024_3dgstream}) and the offline 4DGS~\citep{wu2024_4dgs}, while
    matching their rendering quality. Methods marked with $\dagger$ are reproduced by us in the same environment.}
    \label{fig:teaser}
\end{figure}

The early 2020s have seen rapid development in 3D scene representation and rendering techniques, including neural radiance fields (NeRFs) \citep{mildenhall2020nerf}, 3D Gaussian splatting (3DGS) \citep{kerbl20233d}, and their extensions to 4D representations \citep{wu2024_4dgs}. One natural and important extension is real-time Gaussian streaming of free-viewpoint videos (FVVs). These videos provide a time-varying 3D scene that can be observed from any viewpoint, at any time point. Effective capture, storage, and transmission of FVV can serve as invaluable tools for a diverse range of applications, including surgical planning, immersive 3D video conferencing, and interactive media playback. However, achieving sustainable online FVV streaming remains challenging due to the massive storage demands of detailed scene representations and the need for fast reconstruction and rendering speeds. Importantly, representations must be compact for efficient transmission over limited-bandwidth channels. Existing compact FVV methods generally pursue two main directions:

\textbf{Rate-aware encoding:} QUEEN~\citep{girish2024queen} compresses consecutive Gaussian residuals through quantized latent decoding and sparse positional updates. While quantization of non-position attribute residuals gives substantial storage savings, motion remains challenging to quantize at a per-Gaussian level. 4DGC~\citep{hu2025_4dgc} uses a compact motion grid and sparse compensated Gaussians. Motion grid size is efficiently reduced through joint rate-distortion optimization, but it still remains the dominant component of the inter-frame payload.

\textbf{Anchoring parameterization:} iFVC~\citep{tang2025ifvc} combines anchors with a binary transformation cache and rate-distortion optimization; ReCon-GS~\citep{fu2025recongs} utilizes a density-adaptive multi-level anchor hierarchy; HiCoM~\citep{gao2024hicom} models coherent motion hierarchically; and ComGS~\citep{chen2025comgs} propagates sparse keypoint motion through local influence fields. By sharing transformations across spatially correlated Gaussians, these methods notably reduce the number of independent motion parameters. However, anchor attribute residuals in the shared hierarchical field still bring high storage costs. This leaves direct low-precision coding of small anchor residuals as a distinct compression opportunity.

Here, we propose \textbf{Qu}antized \textbf{A}nchored \textbf{R}esidual \textbf{C}oding \textbf{G}aussian \textbf{S}platting (QuARC-GS), a hierarchical anchor-based online reconstruction framework that substantially reduces per-frame storage while preserving reconstruction fidelity and rendering speed. QuARC-GS introduces \textit{Quantized Hierarchical Anchor Deformation}, which applies straight-through estimator (STE) \citep{bengio2013ste} quantization to hierarchical anchor residuals in the forward pass, effectively eliminating static anchors' transmission costs while compressing dynamic anchors. Furthermore, to better maintain compressed storage for long video streaming, QuARC-GS also features a complementary \textit{Change-gated Densification} module, which creates new Gaussians only where frame-to-frame appearance changes remain after anchor motion has been applied.

Extensive experiments on N3DV and MeetRoom demonstrate improved storage--quality--speed trade-offs, reducing per-frame storage by up to $11\times$ while maintaining competitive reconstruction quality and training speed. Our contributions are summarized as follows:
\begin{itemize}
    \item \textbf{QuARC-GS:} We present a compact online FVV reconstruction framework that combines a persistent multi-level anchor motion representation with compact per-frame appearance updates.
    \item \textbf{Quantization-aware anchor residuals:} We optimize per-anchor motion residuals using their forward-pass quantized values, so the deformation rendered during training matches the transmitted motion payload.
    \item \textbf{Change-gated densification:} We use temporal change after anchor deformation to gate incremental Gaussian creation, reducing redundant appearance capacity in static but difficult regions.

\end{itemize}

\section{Related Work}
\label{sec:related}

\subsection{Dynamic Scene Representations}

Dynamic novel-view synthesis (NVS) aims to reconstruct time-varying scenes from multi-view observations, and render them from unseen viewpoints \citep{li2022neural3dvideo}. The development of neural radiance field (NeRF) methods \citep{mildenhall2020nerf} enabled high-quality NVS through neural representations. Subsequent work made leaps and bounds in their rendering quality, training efficiency, and representation compactness \citep{barron2022mipnerf360, muller2022instantngp, chen2022tensorf}. Dynamic NeRF methods model temporal change through time-conditioned latent representations, canonical deformation fields, or factorized spatiotemporal features \citep{attal2023hyperreel, pumarola2021dnerf, fridovichkeil2023kplanes}.

More recently, 3D Gaussian Splatting (3DGS) \citep{kerbl20233d} exploded in popularity, introducing an explicit anisotropic-Gaussian representation with differentiable rasterization for real-time, high-fidelity rendering. For dynamic scenes, an effective technique is deformation over a set of Gaussians: Dynamic 3D Gaussians \citep{luiten2024dynamic} uses constrained motion to achieve consistent tracking, while deformable 3DGS \citep{yang2024deformable} deforms a canonical Gaussian set. Along the same lines, Geometry-Aware Deformable GS \citep{lu2024gdgs} adds geometric priors, and Grid4D \citep{xu2024grid4d} decomposes input dimensions and applies attention to predict deformations. Complementarily, spatiotemporal approaches tie dynamic scene behavior to where and when it is: ST-GS \citep{li2024stgs} temporally parameterizes Gaussian attributes; TGH \citep{xu2024tgh} uses a temporal hierarchy for long sequences. 4DGS \citep{wu2024_4dgs} uses 4D neural voxels and predicted deformations from HexPlane features~\citep{fridovichkeil2023kplanes,cao2023hexplane} and \citep{yang2024_4dgaussians} 
optimizes 4D primitives for 4D Gaussians. Separately, SC-GS \citep{huang2024scgs} and SP-GS \citep{wan2024superpointgs} use sparse shared motion controls. 

A further line of work targets compactness and efficiency directly: SaRO-GS \citep{yan2024sarogs} models motion with a scale-aware residual field and adaptive optimization for temporally complex scenes; Swift4D \citep{wu2025swift4d} adaptively separates static and dynamic primitives in a divide-and-conquer scheme; and SplineGS \citep{yoon2025splinegs} represents Gaussian trajectories with smooth splines. These representations have achieved notable improvements in computation and rendering speed in offline settings.

\subsection{Compact 3D Gaussian Representations}

Compression of 3DGS representations for efficient streaming and storage is an area of active research. Compressed 3D Gaussian Splatting \citep{niedermayr2024compressed3dgs} applies sensitivity-aware vector clustering and quantization-aware fine-tuning. In contrast, Scaffold-GS \citep{lu2024scaffoldgs} generates local Gaussians from sparse anchors, which HAC \citep{chen2024hac} augments using a hash grid and adaptive quantization. LightGaussian \citep{fan2024lightgaussian} combines importance-based pruning, SH distillation, and VecTree quantization. Beyond reducing model size, methods consider how a Gaussian representation can be transmitted and decoded in a practical system. CodecGS \citep{lee2025codecgs} predicts attributes from progressive tri-planes encoded with standard video codecs, and PCGS \citep{chen2026pcgs} progressively adds anchors while refining quantization precision. A recent plug-and-play approach \citep{bai2026plugandplay} combines feature distribution regularization, opacity-based pruning, and over-compression compensation; QuantizationGS and EntropyGS, respectively, use quantization-aware optimization and attribute-specific entropy models \citep{ma2026quantizationgs, huang2026entropygs}. These methods have made substantial contributions towards a compact static 3DGS model, and allow us to look towards compression in dynamic FVV streaming where memory demands compound.

\subsection{Compact Gaussian Streaming}
The demands on scene representation efficiency become even greater for online FVV, which must compress, transmit, and render a 4D scene from an arbitrary perspective in real time.  To this end, 3DGStream \citep{sun2024_3dgstream} introduces a small hash-based MLP optimized per-frame for Gaussian translation and rotation updates. ReCon-GS \citep{fu2025recongs} represents inter-frame motion using density-adaptive multi-level anchors and dynamically reconfigures the hierarchy to preserve localized motion expressiveness, drawing on many of the past fundamentals of compact 3DGS and serves as strong inspiration for our paper. More recent online methods reduce redundant updates directly: CPOStream \citep{bao2026cpostream} freezes Gaussians predicted to be inactive and recognizes new Gaussians with motion detection. MoRGS \citep{lee2026morgs} uses sparse optical-flow supervision and per-Gaussian motion confidence to down-weight residual updates in static regions.

QUEEN \citep{girish2024queen} is especially relevant to our approach and encodes sequential Gaussian residuals using quantized latent decoding for non-positional attributes and sparse gated positional updates. Motion Matters (ComGS) \citep{chen2025comgs} instead transmits sparse motion-sensitive keypoints, propagates their motion through learned spatial influence fields, and selectively corrects key frames. iFVC \citep{tang2025ifvc} combines anchor Gaussians, a binary transformation cache, and joint rate-distortion optimization, while 4DGC \citep{hu2025_4dgc} jointly optimizes motion-grid compression and sparse compensated Gaussians. Complementary codec-oriented approaches restructure dynamic content as video-coded attributes, motion-layered Gaussian groups, UV atlases, progressive bitstreams, or post-hoc Gaussian-frame codecs \citep{li2025gifstream, wang2024v3, dai2025_4dgv, ke2026streamstgs, rai2026packuv, zheng2025_4dgcpro, zhang2026dfcgs}. These approaches address complementary rate-distortion-latency trade-offs. QuARC-GS builds on these approaches to achieve greater FVV compression without compromising quality.

\begin{figure*}[t]
    \centering
    \includegraphics[width=0.8\textwidth]{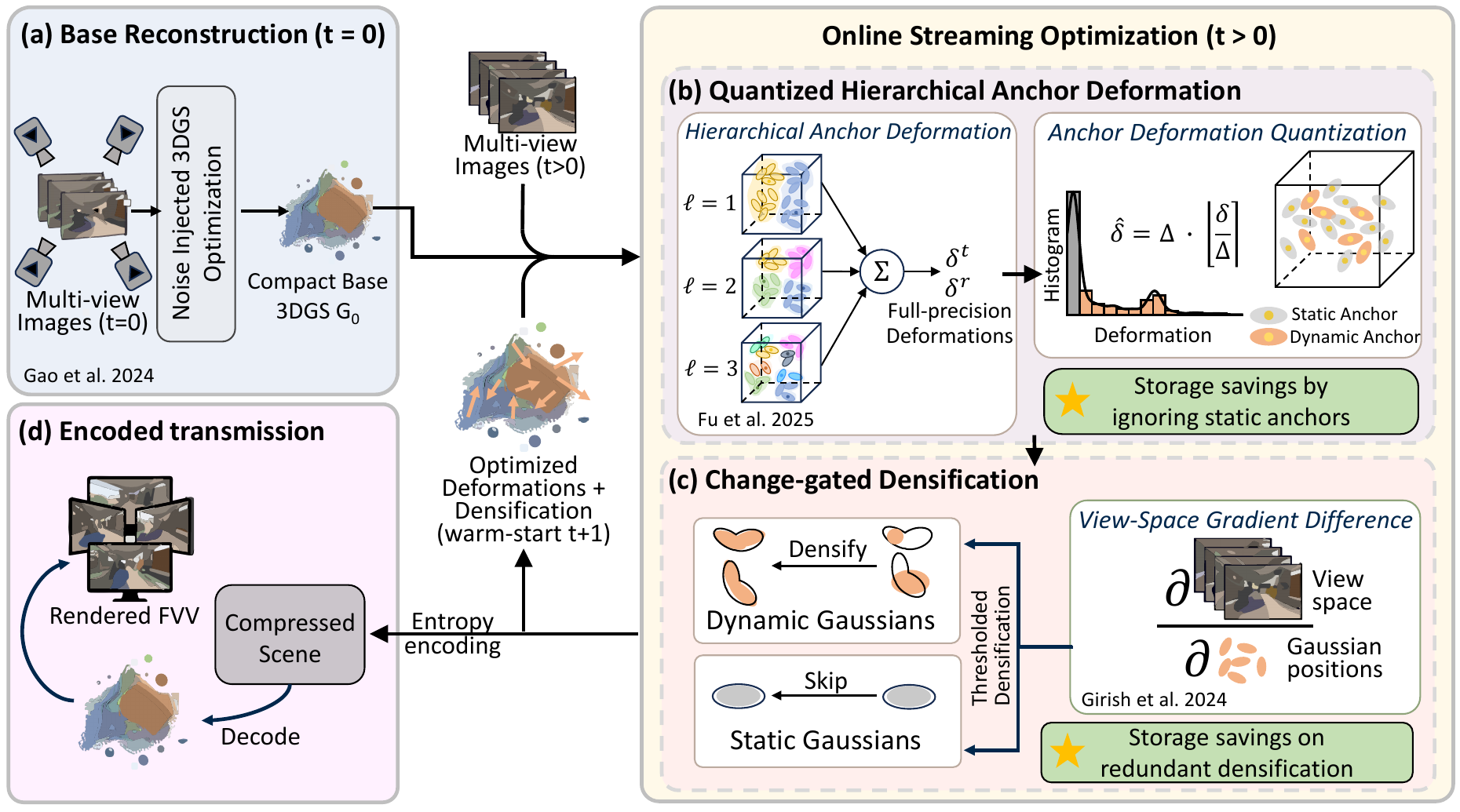} 
    \caption{
    \textbf{Overview of QuARC-GS for online FVV:} (a) QuARC-GS starts with a base 3DGS from multi-view images at $t=0$ optimized with noise injection~\citep{gao2024hicom}. This canonical 3DGS will be causally transformed to represent the scene at subsequent time points ($t>0$). (b) Following \citep{fu2025recongs}, QuARC-GS takes advantage of a hierarchical anchor-based representation of the deformation of the Gaussian 6D poses ($\delta_r, \delta_t$). Our insight was to quantize these deformations, resulting in a large number of static anchors thereby delivering substantial storage savings. (c) Complementarily, QuARC-GS includes a novel module that stems growth in number of Gaussians based on view-space gradients, resulting in storage savings.
    (d) The quantized anchor residuals and incremental Gaussian attributes are used to warm-start the next frame, or entropy-coded and transmitted to enable FVV.
    }
    \label{fig:pipeline}
\end{figure*}

\section{Methods}
\label{sec:method}

\subsection{Preliminaries: 3D Gaussian Splatting}
\label{sec:prelim}
A scene is represented as a sparse collection of anisotropic 3D Gaussians.
Each Gaussian $g_i$ carries a position $\boldsymbol{\mu}_i\!\in\!\mathbb{R}^3$,
a rotation quaternion $\mathbf{q}_i\!\in\!\mathbb{R}^4$, a scale
$\mathbf{s}_i\!\in\!\mathbb{R}^3$, an opacity $\alpha_i\!\in\!\mathbb{R}$,
and spherical-harmonic (SH) color coefficients $\mathbf{c}_i$. Its spatial
extent is described by the covariance
$\boldsymbol{\Sigma}_i=\mathbf{R}_i\mathbf{S}_i\mathbf{S}_i^\top\mathbf{R}_i^\top$,
where $\mathbf{R}_i$ is induced by $\mathbf{q}_i$ and
$\mathbf{S}_i=\operatorname{diag}(\mathbf{s}_i)$. With degree-$1$ SH, each
Gaussian is represented by a $23$-dimensional vector
($3\!+\!4\!+\!3\!+\!1\!+\!3\!+\!9$ for position, rotation, scale,
opacity, base RGB / SH DC, and degree-$1$ SH coefficients). 

For rendering, Gaussians are projected onto the camera plane, depth sorted, and composited by front-to-back alpha blending. A differentiable rasterizer allows attributes to be optimized directly from multi-view images, where 3DGS uses an objective that combines pixel-wise $\mathcal{L}_1$ and D-SSIM losses:
\begin{equation}
    \mathcal{L}_{\mathrm{3DGS}}
    =
    (1-\lambda)\mathcal{L}_1
    +
    \lambda\mathcal{L}_{\mathrm{D\text{-}SSIM}}.
\end{equation}
During optimization, adaptive density control clones, splits, or prunes Gaussians depending on view-space gradients~\citep{kerbl20233d}.

\subsection{Overview}
\label{sec:overview}
To enable practical FVV streaming, the scene must be rapidly reconstructed and compressed into a lightweight representation to facilitate high-speed transmission, while still preserving high-fidelity rendering. Our framework reconstructs photorealistic FVV streams from multi-view video under an online, per-frame optimization paradigm. For the initial frame, we optimize a compact set of base Gaussians, $\mathcal{G}_B$, using noise-injected 3D Gaussian Splatting (3DGS), where small perturbations are applied to Gaussian positions during training to improve robustness~\citep{gao2024hicom}. For each subsequent frame, the optimization is warm-started from the previous frame and models scene motion using adaptive hierarchical-anchor deformation~\citep{fu2025recongs}, which represents deformations through a coarse-to-fine hierarchy of a small number of anchor Gaussians, with each deformation level governed by its own anchor set. Despite this hierarchical representation, the per-frame anchor transformations and newly spawned Gaussians still exhibit substantial temporal redundancy.

To capitalize on this redundancy, we introduce two complementary modules. First, \emph{quantization-aware anchor deformation} incorporates anchor transform quantization directly into the optimization loop using an STE~\citep{bengio2013ste}, enabling the model to optimize the same quantized values that are transmitted while allowing gradients to compensate for quantization errors with minimal quality degradation. Second, \emph{change-gated densification} identifies motion-sensitive Gaussians using the view-space gradient of the rendering-loss difference between consecutive frames and restricts densification accordingly, ensuring that new Gaussians are introduced only in genuinely dynamic regions while static content is not redundantly re-encoded.

\subsection{Hierarchical Anchor Deformation}
\label{sec:anchor}
Optimizing and transmitting a per-Gaussian motion vector for every frame is prohibitively expensive, while a sparse single-level anchor representation lacks the capacity to model complex scene dynamics. Following \citep{fu2025recongs}, we instead represent scene motion with a compact hierarchy of \emph{anchors}, allowing each anchor to share its deformation across the Gaussians bound to it. We employ a hierarchy of $L\!=\!3$ levels. At the finest level, $A_1=\lceil N / m \rceil$ anchors are initialized, where $N$ is the total number of Gaussians and $m$ denotes the maximum number of Gaussians assigned to each anchor. Each subsequent level $\ell$ has $A_{\ell}=\lfloor A_{\ell-1}/r\rfloor$ anchors, with $r=3$ the anchor count ratio between adjacent levels. Thus there are a total of $A=\sum_{\ell=1}^{L} A_{\ell}$ anchors. Anchor positions are generated by uniformly subsampling the current Gaussian positions on a voxel grid. Each Gaussian is then associated with its nearest anchor at every hierarchical level through a $1$-NN search, producing the binding map $\mathcal{I}\in\{1,\dots,A\}^{L\times N}$, which governs the Gaussians' deformations.

Each anchor $a$ stores a deformation parameter $\delta_a=(\delta_a^{t},\delta_a^{r})\in\mathbb{R}^{7}$, consisting of a translation $\delta_a^{t}\!\in\!\mathbb{R}^3$ and a rotation residual $\delta_a^{r}\!\in\!\mathbb{R}^4$ defined with respect to the identity quaternion. The transformed position and rotation of Gaussian $i$ are obtained by aggregating the deformations of its assigned anchors across all hierarchy levels:
\begin{equation}
    \label{eq:deform}
    \mu_i' = \mu_i + \!\!\sum_{\ell=1}^{L}\!\delta^{t}_{\mathcal{I}[\ell,i]},
    \qquad
    q_i' = \mathrm{normalize}\!\Big( \!\sum_{\ell=1}^{L}\!\delta^{r}_{\mathcal{I}[\ell,i]} \Big).
\end{equation}
Consequently, only the compact anchor deformation field $D=\{\delta_a\}\in\mathbb{R}^{A\times 7}$ is optimized and transmitted, substantially reducing the amount of motion data compared to a per-Gaussian deformation representation.

\paragraph{Zero-cost binding.} A key property is that the anchor sampling is a \emph{deterministic, seeded} function of the canonical geometry and the frame index $t$: the voxel subsampling and the $1$-NN binding are re-derived from a pseudo-random seed keyed by $(t,\ell)$. Since the canonical geometry and $t$ already fix the anchor set and binding map $\mathcal{I}$, the binding need not be stored and contributes \emph{zero bytes}; only $D$ is stored.

\subsection{Quantization-Aware Anchor Deformation}
\label{sec:motion}
Even in a compact hierarchical anchor representation, the deformation field is highly redundant: most anchors barely move between frames. Thus, optimizing $D$ in full precision and rounding it afterward is wasteful: small, imperceptible motions consume bitrate, and post-hoc rounding introduces a train/transmit mismatch that degrades reconstruction. To address this, we adopt a quantization framework during training~\citep{girish2024eagles,girish2024queen}. Each attribute of the anchor residual is rounded to a fixed step during the forward pass with an STE, which rounds in the forward pass and passes the gradient through unchanged~\citep{bengio2013ste}:
\begin{equation}
    \label{eq:ste}
    \hat{\delta}^{t}_a = \Delta_{t}\,\big\lfloor \delta^{t}_a / \Delta_{t} \big\rceil,
    \quad
    \hat{\delta}^{r}_a = \mathbf{e} + \Delta_{r}\,\big\lfloor (\delta^{r}_a-\mathbf{e}) / \Delta_{r} \big\rceil,
\end{equation}
where $\lfloor\cdot\rceil$ is rounding, $\mathbf{e}=(1,0,0,0)$ is the identity quaternion, and $\Delta_t $ and  $\Delta_r$ are the translation and rotation quantization steps, respectively. While the rounded, transmitted value $\hat{D}$ is used to transform the Gaussians via Eq.~\eqref{eq:deform}, the gradient still flows through the full-precision $D$ in the backward pass, so that surviving anchors adapt to the quantization.

\paragraph{Static and dynamic sparsity.} The proposed quantization naturally induces a dynamic/static segmentation of the deformation field. Anchor residuals whose magnitude is below $\Delta/2$ are mapped to the identity transform after rounding, whereas larger residuals survive as non-zero motion parameters. Consequently, the anchors are partitioned into a \emph{dynamic} set, containing at least one non-identity quantized residual, and a \emph{static} set, whose residuals all quantize to the identity. The quantized deformation field $\hat{D}$ is entropy-coded independently for translational and rotational components: after quantization, both are represented by integer-valued symbols, and the coding cost is estimated using its empirical Shannon entropy. Since static anchors are mapped to the identity transform, they accumulate in the zero bin, increasing its probability mass and lowering the overall coding cost. As a result, sparsity is captured directly by the entropy coder, without requiring any separate mask or gating term to transmit.

\subsection{Change-Gated Densification}
\label{sec:gate}
Hierarchical deformation alone cannot model changes such as disocclusions, newly visible regions, or appearance variations caused by lighting changes. These effects require incremental Gaussians to model per-frame appearance residuals. However, indiscriminate densification unnecessarily increases the representation size by re-encoding static content. To address this, we propose a change-gated densification strategy that leverages view-space gradients to identify regions containing non-negligible temporal changes ~\citep{girish2024queen} to restrict additions of new Gaussians. In particular, after applying the anchor deformation (Eq.~\eqref{eq:deform}), we compute, over a subset of training views, the screen-space gradient magnitude of the difference between the photometric losses with respect to the current and previous frames,
\begin{equation}
    \label{eq:gate}
    g_i = \Big\| \nabla_{\!\mathbf{x}_i}\!\big[\, \mathcal{L}_1(\hat{I},\,I_t) - \mathcal{L}_1(\hat{I},\,I_{t-1}) \,\big] \Big\|_2,
\end{equation}
where $\mathcal{L}_1$ is the mean absolute error, $\|\cdot\|_2$ is the L2 norm, $\mathbf{x}_i$ denotes the view-space position of Gaussian $i$, $\hat{I}$ is the rendered image, and $I_t$ and $I_{t-1}$ are the target images at the current and previous frames, respectively. Regions whose appearance is well explained by deformation produce similar reconstruction errors across consecutive frames, resulting in small gradients. In contrast, regions undergoing genuine temporal changes produce large gradients. Since the gradients are evaluated \emph{after} deformation, they isolate only the appearance variations that cannot be explained by geometric motion.

The resulting gradient magnitude is normalized as
\begin{equation}
    \label{eq:gatenorm}
    w_i = \frac{g_i}{g_i + \operatorname{median}(g)},
\end{equation}
where $g_i$ denotes the accumulated gradient magnitude. The normalized weight is then used to modulate the accumulated densification gradient before the adaptive clone-and-split operation, restricting densification to regions exhibiting significant temporal change.

\section{Experiments}
\label{sec:exp}

\begin{table*}[t]
    \centering
    \setlength{\tabcolsep}{1.4mm}
    \begin{tabular}{ll|cccccc}
        \toprule
        Category & Method
        & PSNR (dB)$\uparrow$
        & SSIM$\uparrow$
        & LPIPS$\downarrow$
        & Storage (KB)$\downarrow$
        & Train (sec)$\downarrow$
        & Render (FPS)$\uparrow$ \\
        \midrule
        \multirow{4}{*}{Offline}
        & STG
        & 32.05 & 0.948 & -- & 670 & 20 & 140 \\
        & SaRO-GS
        & 32.15 & -- & -- & 1000 & -- & 40 \\
        & Swift4D
        & 32.23 & -- & -- & 400 & 5.0 & 125 \\
        & SplineGS
        & 32.60 & -- & -- & -- & 11 & 76 \\
        \midrule
        \multirow{10}{*}{Online}
        & Dynamic 3DGS
        & 30.67 & -- & -- & --/9200 & 560 & -- \\
        & StreamRF
        & 30.68 & 0.930 & -- & 17700/31500 & 15 & 12 \\
        & 3DGStream$^\dagger$
        & 31.35 & \cb 0.948 & \cb 0.130 & 7600/7800 & 8.1 & 245 \\
        & 4DGC
        & 31.58 & 0.943 & -- & --/500 & 50 & 168 \\
        & QUEEN-1
        & \ca 32.19 & \cc 0.946 & 0.136 & --/750 & \cc 7.9 & \cc 248 \\
        & HiCoM
        & 31.17 & -- & -- & 900 & 11 & \cb 260 \\
        & ComGS-s
        & 31.87 & 0.943 & \cc 0.132 & \cb 49 & 37 & 91 \\
        & ComGS-l
        & \cc 32.12 & 0.945 & \ca 0.129 & \cc 106 & 43 & 147 \\
        & ReCon-GS$^\dagger$
        & \cb 32.16 & \ca 0.951 & \ca 0.129 & 400/440 & \ca 4.7 & \ca 272 \\
        \cmidrule(lr){2-8}
        & \textbf{QuARC-GS}
        & \cb 32.16 & \ca 0.951 & \ca 0.129 & \ca 36/56 & \cb 5.0 & 243 \\
        \bottomrule
    \end{tabular}%
    \caption{Quantitative comparison on the N3DV dataset. The storage metric
    is reported without and with the initial frame, separated by ``/''; the
    training-time metric includes first-frame training. Methods marked with
    $\dagger$ are reproduced by us with official code in the same environment as a mean of 3 runs. The top-3 results are highlighted in each column.}
    \label{tab:main}
\end{table*}

\begin{table}[t]
    \centering
    \setlength{\tabcolsep}{1.55mm}
    \begin{tabular}{l|ccccc}
        \toprule
        Method
        & \shortstack{PSNR\\(dB)$\uparrow$}
        & \shortstack{SSIM\\$\uparrow$}
        & \shortstack{Storage\\(KB)$\downarrow$}
        & \shortstack{Train\\(sec)$\downarrow$}
        & \shortstack{FPS\\$\uparrow$} \\
        \midrule
        3DGStream$^\dagger$
        & 29.30 & 0.948 & 4000/4100 & 4.77 & 260 \\
        HiCoM
        & 26.73 & -- & 600 & 9.8 & 258 \\
        ReCon-GS$^\dagger$
        & \textbf{30.68} & \textbf{0.949} & 300/390 & \textbf{3.2} & \textbf{303} \\
        \cmidrule(lr){1-6}
        \textbf{QuARC-GS}
        & 30.67 & \textbf{0.949} & \textbf{18/26} & 3.3 & 271 \\
        \bottomrule
    \end{tabular}%
    \caption{Quantitative comparison on the MeetRoom dataset. The storage metric
    is reported without and with the initial frame, separated by ``/''; the
    training-time metric includes first-frame training. Methods marked with
    $\dagger$ are reproduced by us with official code in the same environment.}
    \label{tab:meetroom}
\end{table}

\subsection{Datasets}
\label{sec:data}
We evaluate QuARC-GS on two standard FVV multi-view dynamic-scene datasets:
\begin{itemize}
    \item The Neural 3D Video dataset (N3DV)~\citep{li2022neural3dvideo} comprises
    six indoor dynamic scenes. Each scene contains 18 to 21 synchronized multi-view videos,
    with each video spanning 300 frames captured at 30~FPS and a resolution of
    $2704\times2028$.
    \item The Meeting Room dataset~\citep{li2022streamrf} comprises three dynamic
    scenes across diverse real-world scenarios. Each scene contains 13 synchronized multi-view
    videos, with each video spanning 300 frames captured at 30~FPS and a resolution of
    $1280\times720$.
\end{itemize}
The first view of each multi-view video is held out for testing.

\subsection{Implementation Details}
\label{sec:impl}
\paragraph{Training and Optimization.}
QuARC-GS augments the ReCon-GS anchor-deformation framework~\citep{fu2025recongs}. For Gaussian anchor assignment, we used the default hyperparameters from ReCon-GS. For each scene, a canonical Gaussian representation is first optimized from the initial multi-view observations for $15{,}000$ iterations using standard 3D Gaussian Splatting, with the maximum spherical harmonics (SH) degree set to 1 and a noise-injection coefficient of $0.01$~\citep{gao2024hicom}. Subsequent frames are reconstructed in a strictly causal online manner by initializing from the previous frame and optimizing only the frame-specific residuals, consisting of 100 iterations of quantized deformation followed by 100 iterations of change-gated densification. The canonical Gaussians are deformed through a 3-level density-adaptive hierarchical anchor field with at most 4 Gaussians per grid cell, and all learnable parameters are optimized using Adam. All experiments are conducted using PyTorch on a machine running Ubuntu 24.04 with an NVIDIA RTX PRO 6000 GPU.

\paragraph{Residual Compression and Densification.}
Anchor motion residuals are quantized using an STE with fixed world-space quantization step sizes of $\Delta_t = 0.0017$ for translations and $\Delta_r = 0.01$ for quaternion residuals. This quantization maps negligible residuals to the identity transformation, eliminating bitrate overhead for static anchors while preserving dynamic motion. To control representation growth during long streaming sequences, we employ change-gated densification, which introduces new appearance Gaussians only in regions exhibiting significant view-space photometric changes. The gating criterion is computed from frame-to-frame image gradients over a fixed subset of four training views. 

\begin{figure*}[t]
    \centering
    \includegraphics[width=.95\textwidth]{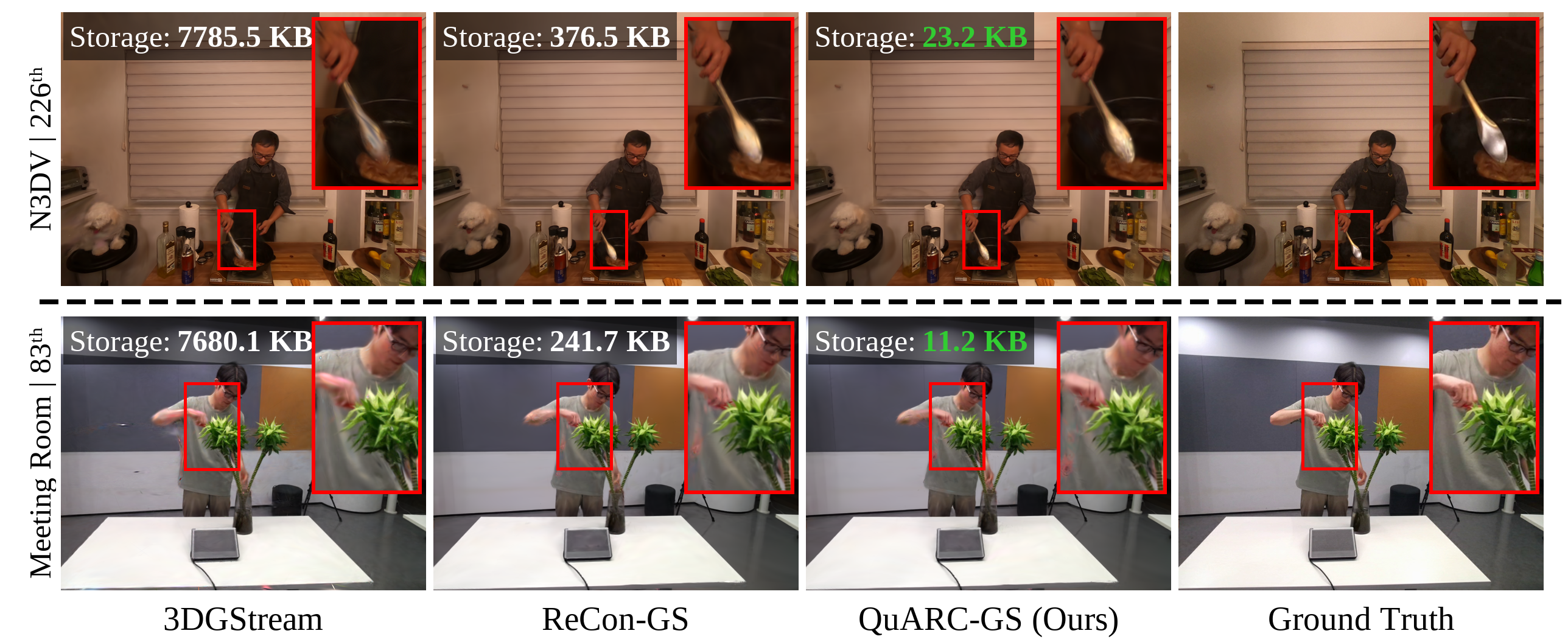}
    \caption{Qualitative comparison of held-out-view renders from QuARC-GS (ours),
    ReCon-GS~\citep{fu2025recongs}, and 3DGStream~\citep{sun2024_3dgstream}
    on the Sear Steak scene from the N3DV dataset~\citep{li2022neural3dvideo} and the trimming scene from the Meeting Room dataset~\citep{li2022streamrf}. QuARC-GS maintains quality while substantially reducing storage.}
    \label{fig:qual}
\end{figure*}

\begin{table}[!h]
    \centering
    \setlength{\tabcolsep}{1.65mm}
    \begin{tabular}{@{}l|ccccc@{}}
        \toprule
        Variant
        & \shortstack{PSNR\\(dB)$\uparrow$}
        & \shortstack{SSIM\\$\uparrow$}
        & \shortstack{LPIPS\\$\downarrow$}
        & \shortstack{Storage\\(KB/f)$\downarrow$}
        & \shortstack{New\\GS/f$\downarrow$} \\
        \midrule
        \textbf{Full (Ours)}
        & \textbf{32.16} & \textbf{0.951} & \textbf{0.129} & \textbf{36/56} & \textbf{153} \\
        w/o quantization
        & \textbf{32.16} & \textbf{0.951} & \textbf{0.129} & 402/441 & 154 \\
        w/o change gate
        & 32.09 & \textbf{0.951} & \textbf{0.129} & 40/60 & 468 \\
        \bottomrule
    \end{tabular}
    \caption{Ablation of the QuARC-GS components on N3DV. ``New GS/f'' is the incremental Gaussians
    added per frame (excluding pruning). The storage metric is reported without and with the initial
frame, separated by “/”.}
    \label{tab:ablation}
\end{table}

\subsection{Quantitative Comparison}
\label{sec:sota}

We evaluate QuARC-GS on N3DV (Table~\ref{tab:main}) and MeetRoom
(Table~\ref{tab:meetroom}) against state-of-the-art (SOTA) offline and online dynamic
scene reconstruction methods. As shown in both tables, QuARC-GS matches the rendering
quality of the best online SOTA in terms of PSNR, SSIM, and LPIPS (i.e., standard metrics for NVS rendering quality) while transmitting
only $36$~KB per frame on N3DV and $18$~KB per frame on MeetRoom. This represents only
a small fraction of the payload required by even the most compact SOTA, yet incurs no
measurable loss in fidelity. These storage savings arise because quantization-aware anchor
deformation collapses static motion to the identity transformation, while change-gated
densification suppresses redundant Gaussians, allowing the transmitted bitstream to
encode only genuinely dynamic content.

Crucially, this compact representation does not come at the cost of speed. QuARC-GS
trains and renders nearly as quickly as the fastest online SOTA while maintaining real-time performance. In sum, QuARC-GS substantially reduces streaming storage requirements
while preserving both the rendering quality and runtime performance of SOTA
online methods.

\subsection{Qualitative Comparison}
\label{sec:qual}
Although QuARC-GS is designed to aggressively compress the per-frame streaming payload,
its rendering quality remains on par with state-of-the-art online methods, as reflected by
the quantitative metrics in Tables~\ref{tab:main} and~\ref{tab:meetroom}. Figure~\ref{fig:qual}
compares held-out-view renders from QuARC-GS against ReCon-GS~\citep{fu2025recongs} and
3DGStream~\citep{sun2024_3dgstream}. Our reconstructions preserve fine texture, dynamic
object boundaries, and thin structures without noticeable blur or artifacts, showing that the
order-of-magnitude reduction in storage does not come at a perceptible cost in visual fidelity.

\subsection{Ablation Studies}
\label{sec:ablation}
\begin{figure}[t]
    \centering
    \includegraphics[width=\linewidth]{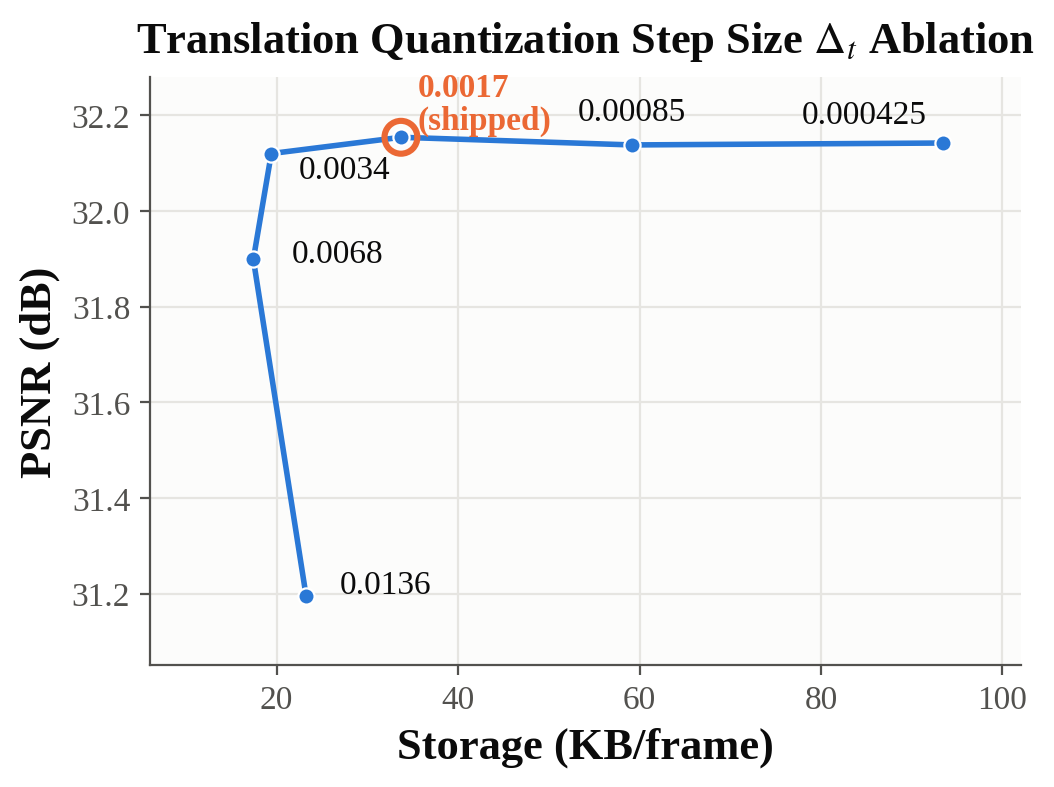}
    \caption{Rate-distortion frontier (PSNR versus per-frame storage on N3DV)
    as the anchor quantization step ($\Delta_t$) varies. The shipped configuration lies at the
    knee of the frontier.}
    \label{fig:quantsweep}
\end{figure}

We ablate the two core components of QuARC-GS on N3DV (Table~\ref{tab:ablation}).
Disabling quantization-aware anchor deformation leaves the reconstruction quality
essentially unchanged but substantially increases the per-frame storage. Without
quantization, every anchor must transmit a full-precision motion residual. In contrast,
quantization snaps residuals onto a shared world-space grid and collapses near-static
anchors to the identity transformation, producing a significantly more compact motion
representation for entropy coding. The quantization step controls this trade-off, as
illustrated in Figure~\ref{fig:quantsweep}: refining the step yields negligible quality
gains while rapidly increasing storage, whereas coarsening it beyond a certain point
degrades reconstruction quality without further reducing storage, since overly coarse
motion becomes misfitting and triggers additional densification. Our chosen step lies at the
knee of this rate--distortion curve, attaining the best quality at minimal storage.

\begin{figure}[t]
    \centering
    \includegraphics[width=\linewidth]{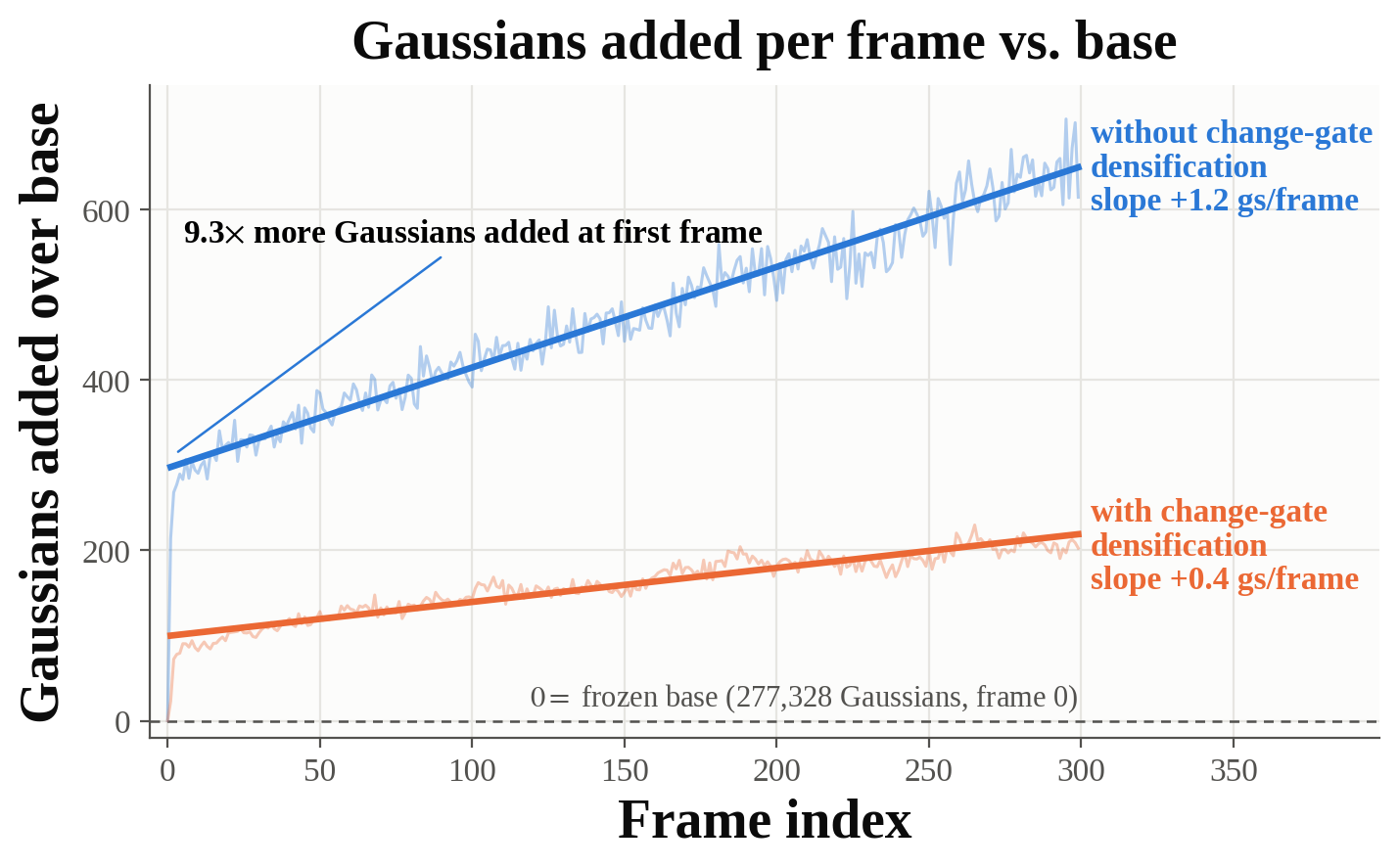}
    \caption{Net total Gaussians added per frame on N3DV, with and without change-gated
    densification. The change gate confines new Gaussians to genuinely changing
    regions, keeping the scene size nearly constant over the sequence.}
    \label{fig:gateablation}
\end{figure}

Disabling the change gate likewise preserves reconstruction quality but substantially
increases the number of Gaussians added per frame, as conventional densification also expands static, 
high-detail regions. As shown in Figure~\ref{fig:gateablation}, this causes the scene size to grow steadily
over time, whereas the proposed change gate restricts densification to genuinely changing
regions, maintaining a nearly constant scene size and a storage-efficient representation 
for long streaming sequences.

Together, quantization-aware anchor deformation reduces the motion payload, while
change-gated densification limits the appearance payload. Both components contribute
substantially to the compactness of QuARC-GS without measurable degradation in
reconstruction quality.

\section{Conclusion}
\label{sec:conclusion}

We presented QuARC-GS, an ultra-compact framework for online FVV reconstruction with Gaussian Splatting. We combine multi-level anchors with entropy-coded, quantization-aware motion, and use thresholded view-space gradients to allocate new Gaussians only to genuinely new or disoccluded content rather than static regions with high reconstruction error. Experiments on multi-view videos demonstrate that QuARC-GS reduces per-frame storage by up to 11$\times$ without compromising quality and speed. 

Although we have substantially improved on storage, there are still limitations of our approach that must be addressed for practical FVV. First, our approach requires a high-quality initial 3DGS with high upfront computational time cost for each video. Further, although the per-frame optimization time of QuARC-GS is competitive with SOTA approaches, it is still too slow for true real-time streaming. Future work on closing this latency gap will complement the storage savings we demonstrated, ultimately making online FVV a reality.

\end{document}